# Rapid and Highly Efficient Synergistic Sonophotocatalytic Degradation of Methyl Orange with Cu-Doped LaFeO$_3$ Perovskite Nanoparticles

Salma Elmouloua [a,b*], M'barek Amjoud [a], Daoud Mezzane [a,b], Manal Benyoussef [b], Jaafar Ghanbaja [c], Mohamed Gouné [d], Mohamed Lahcini [a,e], Zdravko Kutnjak [f], Mimoun El Marssi [b]

[a] *IMED-Lab, Cadi Ayyad University, Av. A. El Khattabi, P.B. 549, Marrakesh 40000, Morocco*

[b] *LPMC, University of Picardy Jules Verne, 33 rue Saint Leu, Amiens 80039, France*

c *IJL, University of Lorraine, CNRS, F-54000, Nancy, France*

[d] *ICMCB, University of Bordeaux, 87 Avenue du Dr Albert Schweitzer, Pessac 33600, France*

[e] *University Mohammed VI Polytechnic, Benguerir 43150, Morocco*

[f] *Advanced Materials Department, Jožef Stefan Institute, Jamova cesta 39, 1000 Ljubljana, Slovenia*

*Corresponding author:*

E-mail: s.elmouloua.ced@uca.ac.ma

ORCID : **0009-0005-9407-9948**

**Abstract**

The integration of sonocatalysis with photocatalysis presents a transformative strategy for advanced wastewater treatment, synergistically overcoming the rapid charge carrier recombination that limits traditional photocatalytic systems. While photocatalysis and sonocatalysis have long been regarded as separate or even competing techniques due to their distinct mechanisms of pollutant degradation, combining them introduces a new paradigm in environmental remediation. In this paper, we show that the combination of sonocatalysis and photocatalysis introduces a new paradigm in advanced wastewater treatment. To this end, we focus on the sonophotocatalytic degradation of methyl orange (MO) using Cu-doped LaFeO₃ perovskite nanoparticles. Remarkably, the Cu-doped LFO nanoparticles exhibited exceptional catalytic performance, achieving a record degradation rate of 0.0455 min⁻¹, which resulted in complete MO removal within 120 minutes when combining sonocatalysis and photocatalysis. A pronounced synergistic effect was demonstrated by the sonophotocatalytic process, as evidenced by a synergy index of approximately 10, underscoring the strong positive interaction between ultrasonic and photonic activation. Robust chemical stability and recyclability were maintained by the engineered nanoparticles, with high sonophotocatalytic efficiency retained over four successive degradation cycles. Hydroxyl radicals (˙OH) and holes (h⁺) were identified as the predominant reactive species responsible for dye degradation through mechanistic investigations via scavenger experiments, and a plausible reaction pathway was proposed. Collectively, the superior sonophotocatalytic activity of Cu-doped LFO relative to its undoped counterpart is highlighted, while the broader potential of dual synergistic sonophotolytic processes for efficient pollutant removal using multiferroic LFO-based materials is illustrated.

**Graphical Abstract**

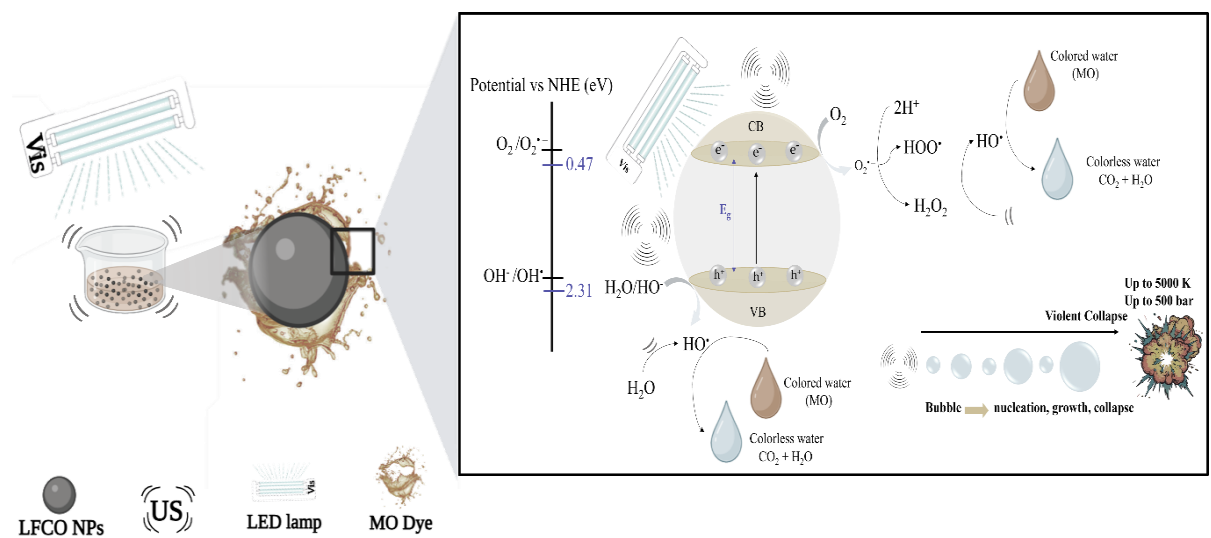

## 1. Introduction

At the international level, the textile industry contributes significantly to environmental pollution, accounting for 8-10% of global carbon emissions[1,2] and up to 20% of industrial wastewater pollution, particularly from azo dyes[1]. These synthetic colorants, comprising 60-70% of all dyestuffs produced[1,2], pose serious threats to aquatic ecosystems and human health due to their non-biodegradability and carcinogenic nature[3]. They have one or more azo groups ($R_1$-N=N-$R_2$) typically with aromatic rings that are substituted mainly by sulfonate groups ($SO_3$), hydroxyl groups ($^{\bullet}OH$), etc.[4,5]. Among azo dyes, Methyl Orange (MO; $C_{14}H_{14}N_3NaO_3S$) is a widely studied anionic dye commonly used in textile applications. Its stable aromatic azo structure and carcinogenic nature raise significant concerns regarding its ecological and human health impacts[6]. Consequently, MO is frequently employed as a model organic pollutant and redox indicator in degradation studies. Traditional water treatment methods such as physical[7], chemical[8], and biological ones[9] used to eliminate dye molecules show several limitations, including low removal efficiencies, the requirement of long removal times, and post-treatments, as well as being costly, and energy-intensive[10]. Advanced Oxidation Processes (AOPs)[11,12] have emerged as a promising solution, utilizing strong oxidizing radical species (e.g., $^{\bullet}OH$ and $^{\bullet}O^{2-}$) to degrade toxic dyes into harmless products such as $CO_2$ and $H_2O$[12]. Among AOPs, photocatalysis[13], sonocatalysis[14,15], and sonophotocatalysis[10,16,17] have gained attention for their efficiency in treating organic dye wastewater. The sonocatalysis technique leverages the physical phenomenon of sonoluminescence[18,19], where ultrasound waves (20 kHz – 2 MHz) induce acoustic cavitation, a dynamic process of microbubble formation, growth, and collapse[10,20]. These microscopic events create intense "hot spots" that concentrate energy and generate reactive radicals, effectively transforming localized microreactors into powerful degradation environments[21]. Simultaneously, photocatalysis utilizes semiconductor catalysts exposed to light to produce reactive oxygen species (ROS) that degrade organic pollutants[22]. However, photocatalysis alone is not suitable for large-scale applications due to its slow degradation response time and several disadvantages of the photocatalyst, including a narrow wavelength response range[24], low solar energy utilization[24], and high-rate electron-hole ($e^−/h^+$) recombination[23,24]. Similarly, sonocatalysis faces challenges in achieving the complete mineralization of pollutants[25]. To address these limitations, researchers have developed a new approach known as "Sonophotocatalysis", which combines sonocatalysis and photocatalysis[18,26]. Sonophotocatalysis, a hybrid approach combining sonocatalysis and photocatalysis, offers enhanced degradation efficiency by synergistically coupling ultrasound and light irradiation. The combination of these techniques in sonophotocatalysis provides several advantages, including greater efficiency in pollutant removal[27], versatility in treating various water contaminants[28], and reduced treatment times[24,29]. Sonophotocatalysis emerges as a groundbreaking solution to these challenges, synergistically combining ultrasonic irradiation with photocatalytic processes to create a robust, self-cleaning catalytic system that dramatically enhances pollutant degradation capabilities[5]. By integrating sonocatalysis and photocatalysis, sonophotocatalysis effectively mitigates issues such as high costs, sluggish activity, and prolonged reaction times. The acoustic cavitation mechanism plays a crucial role in this process, generating transient cavitation bubbles that create localized high-temperature and high-pressure microenvironments, which effectively clean catalyst surfaces and disrupt the molecular structures of pollutants[21]. For instance, Abazari et al.[30] showed that Core@shell Zr(IV) metal-organic framework nanorods and $ZnIn_2S_4$ nanostars (NU@ZIS20) achieved a remarkable sonophotocatalytic activity of 92% for tetracycline degradation after five cycles. This high performance was maintained without significant structural or morphological changes, as confirmed by XRD patterns, and TEM micrographs.

This not only improves the overall efficiency of the treatment process but also potentially increases energy consumption through faster pollutant degradation[29]. Therefore, sonophotocatalysis offers new ideas for environmental improvement research, such as sterilization, organic-pollutant degradation, and water splitting to produce oxygen and hydrogen[31].

Recent research has focused on semiconductor nanomaterials for water treatment due to their exceptional physicochemical properties, nontoxicity, and environmental sustainability[32–34]. Perovskites, with their typical $ABO_3$ structure, where position A is occupied by a rare earth ion, and position B by a transition metal (TM) ion, have emerged as promising catalysts due to their lower band-gap ($E_g$) semiconducting properties[35]. Lanthanum iron oxide $LaFeO_3$ (LFO), one of the most common p-type semiconductors, has garnered significant attention due to its physical, chemical, and magneto-optical properties[13,36]. Due to its multiferroicity, LFO has garnered significant attention in various advanced technology applications, including solid oxide fuel cell catalysts, chemical sensors, catalysis, water splitting, solar cells, and biosensors[37–40]. LFO's structural and chemical stability, abundance, low cost, non-toxicity, and narrow band gap energy (2.0-2.6 eV) make it an attractive material for catalysis applications[41]. Despite its promising properties, LFO faces a significant challenge in its photocatalytic applications due to the high recombination rate of photogenerated ($e^-/h^+$) pairs, which severely limits its catalytic efficiency[41–43]. To address this issue, doping perovskite with metal elements can be a suitable strategy to enhance its charge transport properties and reduce the rate of electron-hole recombination[44]. For instance, transition metal (TM) doping in LFO has shown remarkable improvements in photocatalytic performance[45–47]. The substitution of $Fe^{3+}$ with TM ions (e.g., $Cu^{2+}$, $Mg^{2+}$, $Zn^{2+}$, etc.) not only reduces the band gap, enhancing visible light absorption but also introduces intra-band energy levels that act as electron traps [1,48–50] and introducing oxygen vacancies (OVs), thereby extending the lifetime of photogenerated electrons and holes[51,52]. Among the substitution of TM (Mn, Co, Cu), Peng et al. demonstrated that 10% Cu-doped LFO exhibited a photocurrent density of 0.99 mA/cm$^2$ at 1 V vs. Ag/AgCl, nearly tripling the photocurrent performance of undoped LFO[47]. Electrochemical impedance spectroscopy measurements revealed a significant reduction in charge transport resistance in the doped LFO samples, indicating improved separation and transfer of photogenerated charge carriers[47]. Furthermore, the introduction of Cu$^{2+}$ ions promoted the formation of oxygen vacancies, which serve as active sites for the adsorption and activation of reactant molecules. Moreover, as reported in several studies, Li et al. used an ultrahigh vacuum to create OVs, which became active sites for the generation of hydroxyl radicals on the surface of Bismuth oxychloride (BiOCl)[53]. Wang et al. reported that oxygen vacancies in Cu-doped LaAlO$_3$ were favorable for the dissociation of H$_2$O$_2$ to generate hydroxyl radicals[54]. While these findings highlight the potential of metal ion doping in enhancing LFO's photocatalytic activity, the synergistic effects of combining this approach with sonophotocatalysis remain unexplored, presenting an exciting avenue for future research in AOPs for water treatment.

In this work, pure and 10 at.% Cu-doped LFO nanoparticles (NPs), synthesized using the sol-gel auto-combustion route, were able to harness photonic as well as mechanical energy for the total degradation of Methyl Orange (MO) dye water pollutant. Interestingly, it was observed for the first time that undoped LFO and Cu-doped LFO exhibit a significantly higher photocatalytic efficiency when combined with low-frequency mechanical energy. The structure, morphology, optical characteristics, MO degradation performance, Total Organic Carbon (TOC) measurements, and reactive species trapping experiments of LFO and LFCO NPs were thoroughly characterized.

## 2. Materials and Methods

### 2.1. Synthesis of LFO and LFCO nanopowders

The undoped and Cu-doped LFO NPs were synthesized using the sol-gel auto-combustion method. All starting materials were of analytical grade and used without further purification. Stoichiometric amounts of the nitrate precursor reagents La(NO$_3$)$_3$·6H$_2$O [Alfa Aesar; purity≥ 99.9%], Fe(NO$_3$)$_3$, 9H$_2$O [Oxford; purity ≥ 99.0%], and Cu(NO$_3$)$_2$·3H$_2$O [Acros Organics; purity≥ 99.0%] were dissolved in distilled water under continuous stirring. Thereafter, citric acid was added to the precursor solution as a chelating agent. The molar amount of citric acid (CA) was equal to the total molar amount of metal nitrates in the solution. This mixture was then heated to 80°C with continuous stirring for 1 hour. Subsequently, ethylene glycol (EG) was added as a fuel agent in a molar ratio of CA/EG = 1/4. The solution was further heated to 110°C for 1 hour to achieve gel formation. For the auto-combustion step, the gel was rapidly placed into an electrical oven at 180°C for 1 hour to obtain foamy dry powders. Finally, the obtained powders were annealed at 900°C for 2 hours to remove any remaining organic materials and to form the desired compounds designated as LFO and LFCO.

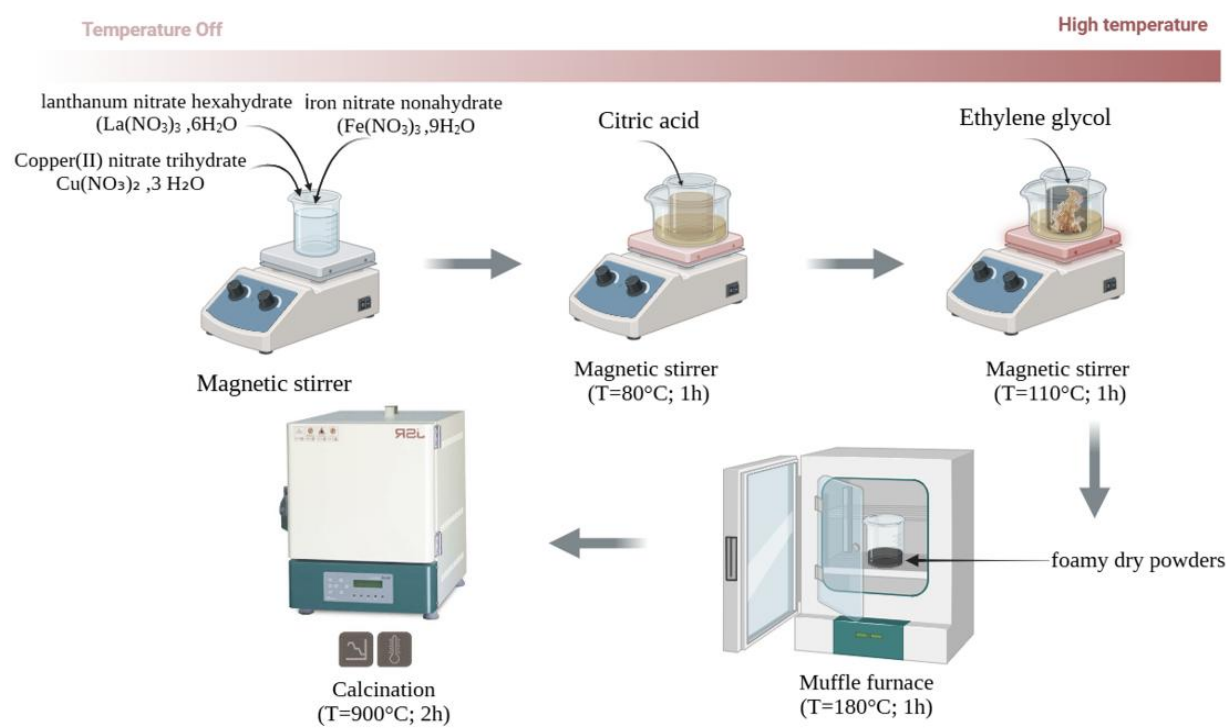

Fig. 1. The synthesis process of LFO and LFCO powder by the sol-gel auto combustion method

## 2.2. Samples characterization

The structure of undoped and Cu-doped LFO nanopowders was characterized at room temperature by X-ray diffraction (XRD) using Cu Kα radiation (λ = 1.54059 Å) on a Rigaku diffractometer. The scanning was performed in the 2θ range from 20° to 80° with a step size of 2° for 1 min per step. Raman spectra of powder samples were recorded at room temperature in the wavenumber range of 100 cm$^{-1}$ – 800 cm$^{-1}$ using a green excitation laser with a Renishaw in Via Reflex Raman spectrometer equipped with an edge filter. Transmission Electron Microscopy (TEM) investigations were conducted using a JEM-ARM 200F Cold FEG TEM/STEM operating at 200 kV and equipped with a spherical aberration (Cs) probe and image correctors. STEM micrographs were acquired using a High-Angle Annular Dark-Field (HAADF) detector. UV–vis diffuse spectra for degradation analysis were measured using a Shimadzu UV-2600 spectrophotometer, with wavelengths ranging from 350 to 650 nm. The zeta potential of the synthesized LFO and LFCO powders was measured at room temperature in an aqueous suspension using a Zetasizer apparatus from Nano ZS, Malvern instruments.

## 2.3. Catalysis Assessment

The degradation of MO dye under visible light illumination and ultrasonic vibration, both individually and simultaneously, was used to evaluate the photocatalytic, sonocatalytic, and sonophotocatalytic efficacy of the synthesized catalysts. LFO and LFCO nanoparticles (10 mg) were dispersed in 10 mL of MO solution (10$^{-5}$ M) in a beaker. The turbid solution with a pH of 6 was mechanically stirred at a speed of 100 rpm in the dark to achieve adsorption-desorption equilibrium before initiating the catalysis evaluation. The dye solution with dispersed catalyst was then exposed to visible light irradiation from a LED lamp (30 W) for the photocatalytic test, to low ultrasonic vibration from an ultrasonicator (37 kHz, 300 W) for the sonocatalytic test, or to both visible light illumination and ultrasonic vibration simultaneously for the sonophotocatalytic test. Subsequently, 2 mL of liquid was collected from the reaction mixture every 30 minutes and centrifuged to obtain the supernatant. The UV-visible absorption of the residual MO in the supernatant was measured to determine the concentration of the dye. The Degradation efficiency was calculated using the following equation (Eq. 1)

$$Degradation\ efficiency\ (\%) = (1 - \frac{C_t}{C_0}) \times 100 \qquad Eq.\ 1$$

Where, $C_0$ is the initial MO concentration (in mg L$^{-1}$) at t = 0 min, and $C_t$ is the MO concentration at time t. To prevent the dye solution from heating up during ultrasonication, the water in the ultrasonicator, which served as the medium for ultrasonic wave propagation, was changed every 10 minutes.

To determine if the sonophotocatalytic degradation of the MO organic dye pollutant led to its partial or total mineralization, the Total Organic Carbon (TOC) corresponding to the remaining organic dye, along with potentially organic byproducts, was measured after the test using a TOC analyzer (TOC-L CPH/CPN, Shimadzu).

During the investigation of the sonophotocatalysis mechanism, appropriate amounts of scavengers were added separately to the dye solution to bind specific active radical species. Isopropanol (IPA) was used to scavenge hydroxyl radicals ($^{\bullet}OH$)[55], circulating N$_2$ was employed to trap the superoxide radicals ($^{\bullet}O^{2-}$)[56], and ethylenediamine tetra-acetic acid disodium (EDTA) was used as a hole (h$^+$) scavenger[55].

## 3. Results and discussions

### 3.1. Microstructural and Structural Analysis

Transmission Electron Microscopy (TEM) was employed to elucidate the morphological and structural characteristics of the synthesized Cu-doped and undoped LFO NPs. Bright Field-TEM (BF-TEM) micrographs revealed a randomly distributed structure with a tendency to form spherical-like shapes and irregular nanoparticles (Fig. 2(a($a_1$)) and b($b_1$))). The Selected Area Electron Diffraction (SAED) patterns confirmed the single-phase orthorhombic structure of both samples. The presence of diffraction spots indicates that the nanoparticles are single crystals, Fig. 2(a($a_2$, $a_3$ ) and b($b_2$, $b_3$ )). SAED analysis demonstrated a well-ordered periodic lattice for both pure LFO and Cu-doped LFO, observed along the zone axis [100]. For undoped LFO, the measured d-spacing was approximately 3.91 Å, which is associated with the (020) plane and closely matches the calculated value of 3.92 Å obtained from XRD data. In the case of Cu-doped LFO, the SAED pattern indicated a d-spacing of 3.53 Å, corresponding to the (111) plane, which is in good agreement with the theoretical value of 3.51 Å. TEM measurements confirmed the random shape of NPs while demonstrating excellent crystalline order within the orthoferrite particles, affirming the formation of high-quality orthorhombic LFO and LFCO NPs. STEM-EDX elemental maps revealed a homogeneous distribution of La, Fe, Cu, and O throughout the sample matrix, as shown in Fig. 2 ((a($a_5$)); b, ($b_5$)). Energy Dispersive X-ray analysis (EDX) (Fig. 2(a($a_4$)) and b($b_4$)) qualitatively confirms the presence of lanthanum, iron, and oxygen in all synthesized NPs. For undoped LFO, the La/Fe atomic ratio was determined to be 1±0.07:1±0.07, consistent with the expected stoichiometry. In LFCO samples, the La/Fe/Cu atomic ratio was 1:0.9±0.01:0.1±0.01, indicating the formation of $LaFe_{0.9}Cu_{0.1}O_3$. Moreover, TEM analysis failed to identify any discrete Cu oxide phases, further substantiating the successful incorporation of Cu into the LFO structure.

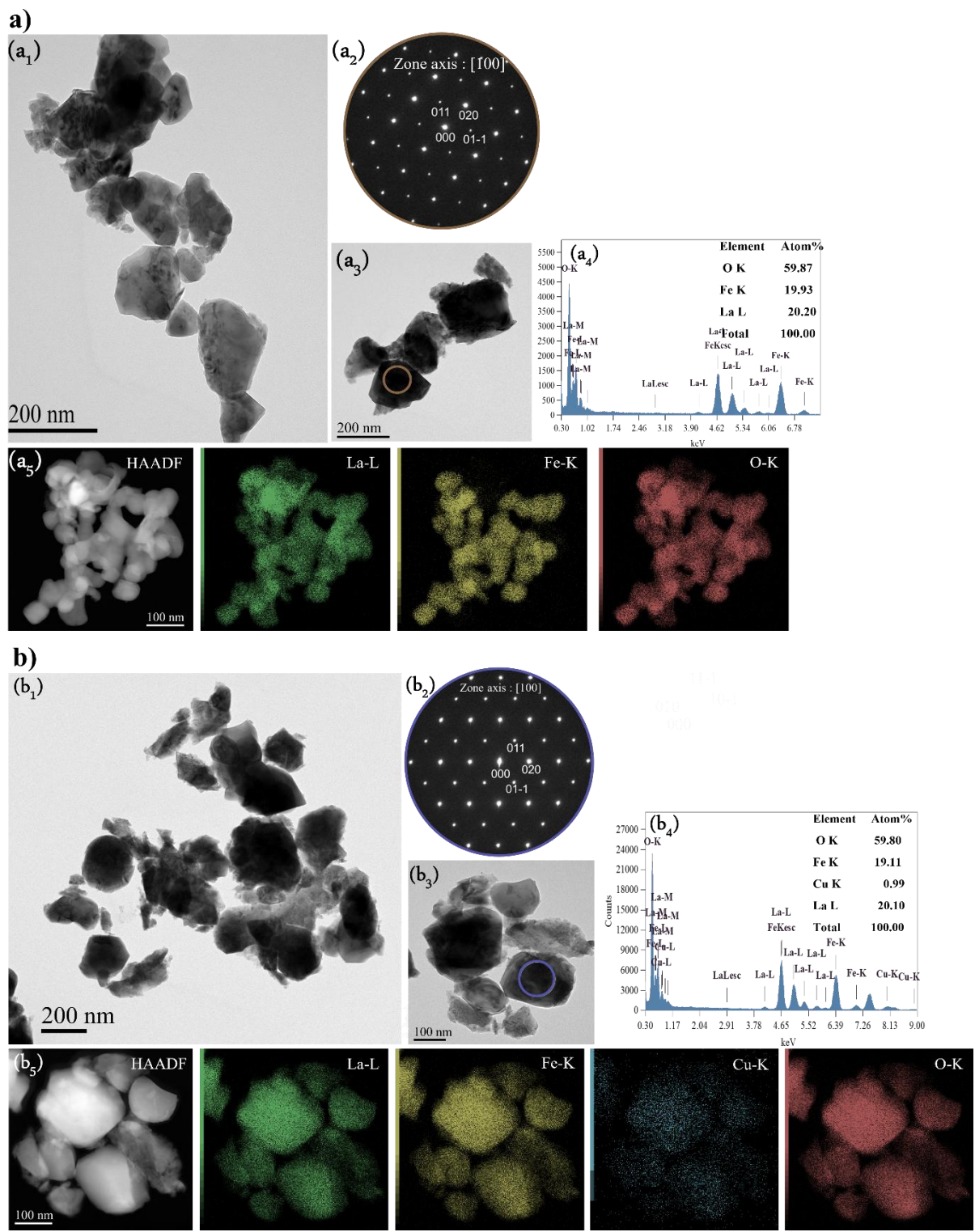

Fig. 2. TEM analysis of (a) LFO NPs, and (b) LFCO NPs:

(a$_1$) BF TEM micrograph, (a$_2$) SAED pattern of particle indicated by the circle in image (a$_3$),

(a$_4$) EDX analysis, (a$_5$) HAADF-STEM and X-elemental maps.

($b_1$) BF TEM micrograph, ($b_2$) SAED pattern of particle indicated by the circle in image ($b_3$), ($b_4$) EDX analysis, ($b_5$) HAADF-STEM and X-elemental maps.

The X-ray diffraction (XRD) analysis at room temperature provided valuable insights into the crystal structure of the LFO and LFCO samples, as shown in Fig. 3. The well-indexed diffraction peaks observed indicate the formation of a single perovskite phase, consistent with previous literature[47,57]. The undoped $LaFeO_3$ sample exhibits a pure orthorhombic perovskite phase (space group Pbnm) [58]. The diffraction patterns of the Cu-doped sample are compared with the reference LFO. It was observed that Cu doping did not alter the orthorhombic structure of LFO, and no copper oxide peaks were detected. Furthermore, Fig. 3(b) highlights an enlarged XRD pattern in the $2\theta$ range of 32-33°, where a minor shift toward higher $2\theta$ angles is induced by the Cu incorporation in stoichiometric LFO. This shift indicates a decrease in the lattice parameter, resulting in a contraction of the unit cell volume, which is corroborated by unit cell parameter calculations, as summarized in Table 1. However, this behavior contrasts with the expected expansion effect, as the ionic radius of $Cu^{2+}$ (0.72 Å) is larger than that of $Fe^{3+}$ (0.64 Å)[59].

The substitution of Cu for Fe in the lattice of $LaFeO_3$ creates a charge imbalance and disturbs the charge neutrality of the system. There are two principal ways to achieve charge compensation for the Cu-substitution: first the presence of oxygen vacancies as in $LaFe_{1-x}Cu_x(O_{3-0.5x}\square_{0.5x})$[60], and the redox reactions in the Fe site. In contrast, the oxygen content is kept constant, as in $La(Fe_{1-x-y}^{3+}Fe_y^{4+}Cu_x^{2+})O_3$ (creation of some smaller Fe cations from (0.64 Å) for $Fe^{3+}$ to (0.58 Å for $Fe^{4+}$)[61]. Similar charge compensation behavior has been reported in related lanthanum-based perovskites such as $LaMn_{1-x}Cu_xO_3$ and $LaCo_{1-x}Cu_xO_3$, which share structural similarities. In $LaMn_{1-x}Cu_xO_3$, increasing Cu content promotes $Mn^{4+}$ formation, reaching complete oxidation (100% $Mn^{4+}$) at x = 0.6. In contrast, $LaCo_{1-x}Cu_xO_3$ contains exclusively trivalent cobalt, with oxygen vacancies serving as the primary charge compensation mechanism for $Cu^{2+}$ substitution. According to various studies in the literature, the predominant mechanism for charge compensation in Cu-doped $LaFeO_3$ perovskites is the formation of oxygen vacancies[61].

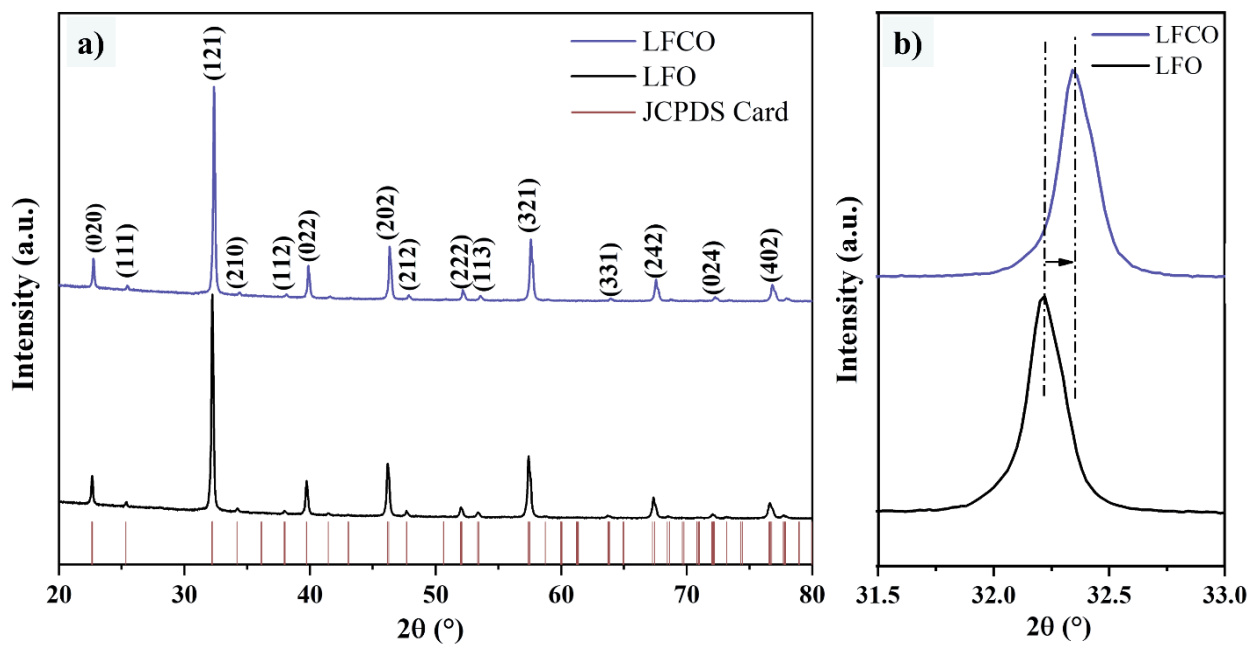

Fig. 3. a) Room temperature X-ray patterns of LFO and LFCO NPs, b) enlarged views of XRD peaks between 31.5 and 33°

Table 1. Lattice parameters, cell volume, and particle size of LFO and LFCO NPs.

| Sample | Lattice parameters | | | Cell volume ($Å^3$) | Atom fraction (at.%) | | |
|---|---|---|---|---|---|---|---|
| | a (Å) | b (Å) | c (Å) | | La | Fe | Cu |
| **LFO** | 5.56 | 7.84 | 5.55 | 242.24 | 20.20 | 19.93 | 0 |
| **LFCO** | 5.54 | 7.83 | 5.54 | 240.12 | 20.10 | 19.11 | 0.99 |

Unlike XRD, Raman spectroscopy offers high sensitivity to local structural distortions and oxygen vibrations in perovskite-type materials. This technique effectively probes the short-range order and bond lengths around $BO_6$ octahedra, providing insights into the material's local structure[56]. The Raman spectrum of pure and Cu-doped LFO NPs, measured in the 30–800 cm$^{-1}$ wavenumber range, reveals distinct vibrational modes. The literature has identified 24 Raman-active modes ($\Gamma = 7A_g + 5B_{1g} + 7B_{2g} + 5B_{3g}$) in rare-earth orthoferrite with an orthorhombic structure (Pbnm space group)[62,63]. The observed Raman modes appear at ~81, ~100, ~136, ~156, ~186, ~266, ~294, ~413, ~433, ~498, and ~627 cm$^{-1}$ (Fig. 4). Below 200 cm$^{-1}$, Raman modes arise from vibrations associated with La site motion[63], while the mode around 186 cm$^{-1}$ corresponds to La rotation[63]. Moreover, the Raman modes at 266, 294, 413, 430, and 498 cm$^{-1}$ originate from ion vibrations within the $Fe^{3+}$ oxygen octahedra [64]. Between 400 and 500 cm$^{-1}$, Oxygen octahedral bending vibrations prevail[63], whereas modes above 500 cm$^{-1}$ correspond to oxygen stretching vibrations[63,64]. Upon Cu doping, two additional peaks emerge at 379 cm$^{-1}$ and 557 cm$^{-1}$, which can be assigned to Cu-O stretching vibrations[65,66]. The absence of copper oxide phases in XRD confirms that these peaks arise from Cu incorporation into the LFO lattice rather than secondary phases. Additionally, one can observe a blue shift upon Cu doping, in addition to a broadening of the modes. This behavior indicates lattice distortion, which can be attributed to the substitution of $Fe^{3+}$ (0.64 Å) by the larger $Cu^{2+}$ (0.73 Å) ions[67]. The incorporation of $Cu^{2+}$ into the $Fe^{3+}$ sites generates oxygen vacancies to maintain charge neutrality due to the difference in oxidation states[66,67]. These vacancies lead to structural modifications, affecting Raman-active modes by disturbing Fe-O-Fe bonds and forming new Cu-O-Fe or Cu-O-Cu bonds[67]. The formation of oxygen vacancies near Cu ions results in lattice contraction, contributing to the observed peak shifts toward higher wavenumbers[67,68]. This phenomenon is consistent with previous studies on similar systems, such as Cu-doped $TiO_2$[67], where the insertion of Cu into the $TiO_2$ lattice weakens the O-Ti-O bonds, leading to changes in Raman frequencies. Additionally, Cu doping reduces the crystallinity of LFO compared to its undoped form. This reduction in crystallinity explains the broader peaks and lower intensities observed in the Raman spectrum of Cu-doped samples[68]. These findings collectively demonstrate how Cu doping alters the vibrational dynamics and structural properties of LFO through lattice distortions and oxygen vacancy formation and consequently their optical and catalytic performances. Based on XRD and Raman spectroscopy results, it can be concluded that $Cu^{2+}$ ions are successfully incorporated into the LFO structure.

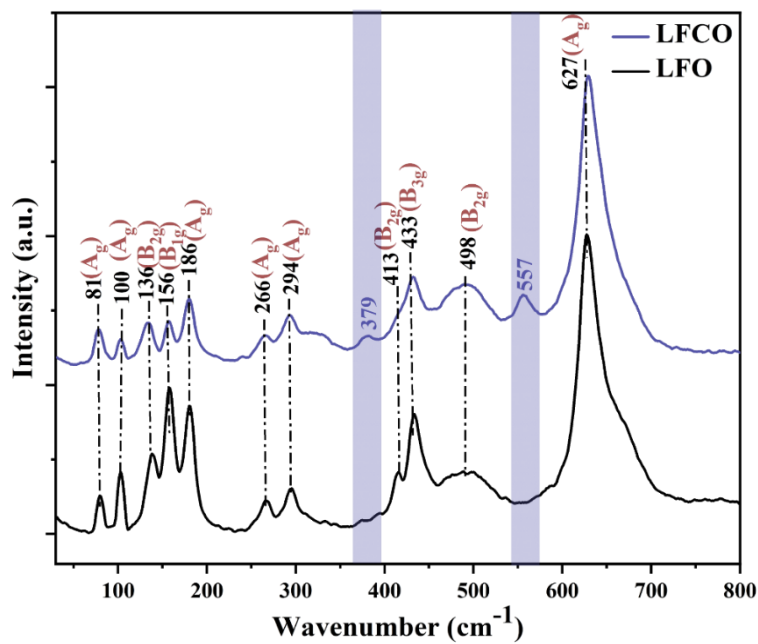

Fig. 4. Raman spectra of LFO and LFCO NPs at Room temperature.

## 3.2. Optical properties

To estimate the energy band gaps of $LaFeO_3$ and Cu-doped $LaFeO_3$ samples, Tauc plots were determined using the Kubelka–Munk method[69]. The Tauc method is based on the assumption that the energy-dependent absorption coefficient (α) can be expressed by the following equations 2 and 3[70]:

$$\alpha = F(R) = \frac{(1-R)^2}{2R} \qquad Eq.\ 2$$

$$(\alpha.h\upsilon)^{\frac{1}{\gamma}} = B(h\upsilon - Eg) \qquad Eq.\ 3$$

Where F(R) is the Kubelka–Munk function; α is the photon absorption coefficient; R is the reflectance, $h$ is the Planck constant, ν is the photon's frequency, $E_g$ is the band gap energy, and B is a constant. The γ factor depends on the nature of the electron transition and is equal to 1/2 or 2 for the direct and indirect transition band gaps, respectively[70]. The optical diffuse reflectance spectra were used to investigate the optical characteristics of the prepared materials and determine their optical band gap. Figure 5b displays the plot (α(hν)$^2$) versus the incident photon energy, fitted with a linear regression, which identifies a direct bandgap transition of 2.3 eV for LaFeO$_3$. This value was consistent with the bandgap range of (2.0–2.6 eV) reported in other studies[11,38,57]. Changes in bandgap energy depend not only on the material type but also on the synthesis method, crystallinity, particle size, and stoichiometry[39]. Based on Figure 5c, the addition of Cu to the LFO structure decreased the direct optical bandgap. The estimated band gap for the Cu-doped LaFeO$_3$ sample is 1.84 eV. Notably, replacing Fe with Cu in the LFO structure modified the light absorption property of perovskite[46]. Meanwhile, Dong et al. reported a reduction in the energy gap with the introduction of $Zn^{2+}$ into LFO[71], suggesting that substituting metal ions at the Fe site in LFO may form oxygen vacancies and additional energy levels, thus increasing atomic distances and narrowing the material's energy band gap[71]. According to the literature[56,57,72], the band edge positions of the catalysts can be calculated using the Mulliken electronegativity. The conduction band (CB) and valence band (VB) potentials of the samples can be calculated using the following equations 4 and 5[55]:

$$E_{VB} = \chi - E^e + 0.5 \times E_g \quad Eq.\ 4$$

$$E_{CB} = E_{VB} - E_g \quad Eq.\ 5$$

$E_{CB}$ and $E_{VB}$ are the conduction and valence band potentials, respectively. $E_g$ the band gap estimated from the UV–vis results. $E^e$ is equal to 4.5 eV and represents the energy of the free electrons versus the hydrogen scale. $\chi$ is the electronegativity of the semiconductor and can be calculated following Eq. 6[72].

$$\chi = [\chi(A)^a \cdot \chi(B)^b \cdot \chi(C)^c \cdot \chi(D)^d \cdot \chi(E)^e]^{\frac{1}{a+b+c+d+e}} \quad Eq.\ 6$$

The parameters a, b, c, d, and e indicate the number of atoms for each element in a composition, with χ(y) representing the electronegativity of each corresponding element. The absolute electronegativity of LaFeO₃ is 5.57 eV, whereas LaFeO₃ doped with 10% Cu has an electronegativity of 5.85 eV on the Pauling scale[73]. The estimated CB edge potential versus NHE for pure LFO is −0.08 eV. Upon doping, the CB edges undergo a positive shift, reaching a value of 0.47 eV for 10%Cu-LFO. Consequently, the VB edge for pure LFO is 2.22 eV. However, upon doping, the VB edge values become positive, reaching a value of 2.31 eV for the LFO-10%Cu sample. Introducing a dopant element creates additional energy levels within the band gap due to O-vacancies, which can narrow the band gap. This modification may have significant effects on the energy harvesting of Cu-doped LFO, as it enables the generation of more electrons and holes, thereby enhancing the catalytic activity. Energy band positions play a crucial role in designing catalysts for electrochemical reactions(Fig. 5d). Specifically, at pH =6, the reduction of oxygen ($O_2$) to superoxide ($^{\bullet}O_2^-$) necessitates an $E_{CB}$ positioned below -0.33 V relative to the normal hydrogen electrode (NHE). Conversely, the oxidation of hydroxyl ions ($OH^-$) to hydroxyl radicals ($^{\bullet}OH$) requires an $E_{VB}$ to be more positive than +1.99 V (vs. NHE). Consequently, the energy band positions indicate that Cu-doped LFO would exhibit enhanced efficiency in degrading methyl orange dye by preferentially generating hydroxyl radicals (•OH) over superoxide (•O$_2^-$), due to its optimized valence band alignment (Fig. 5d).

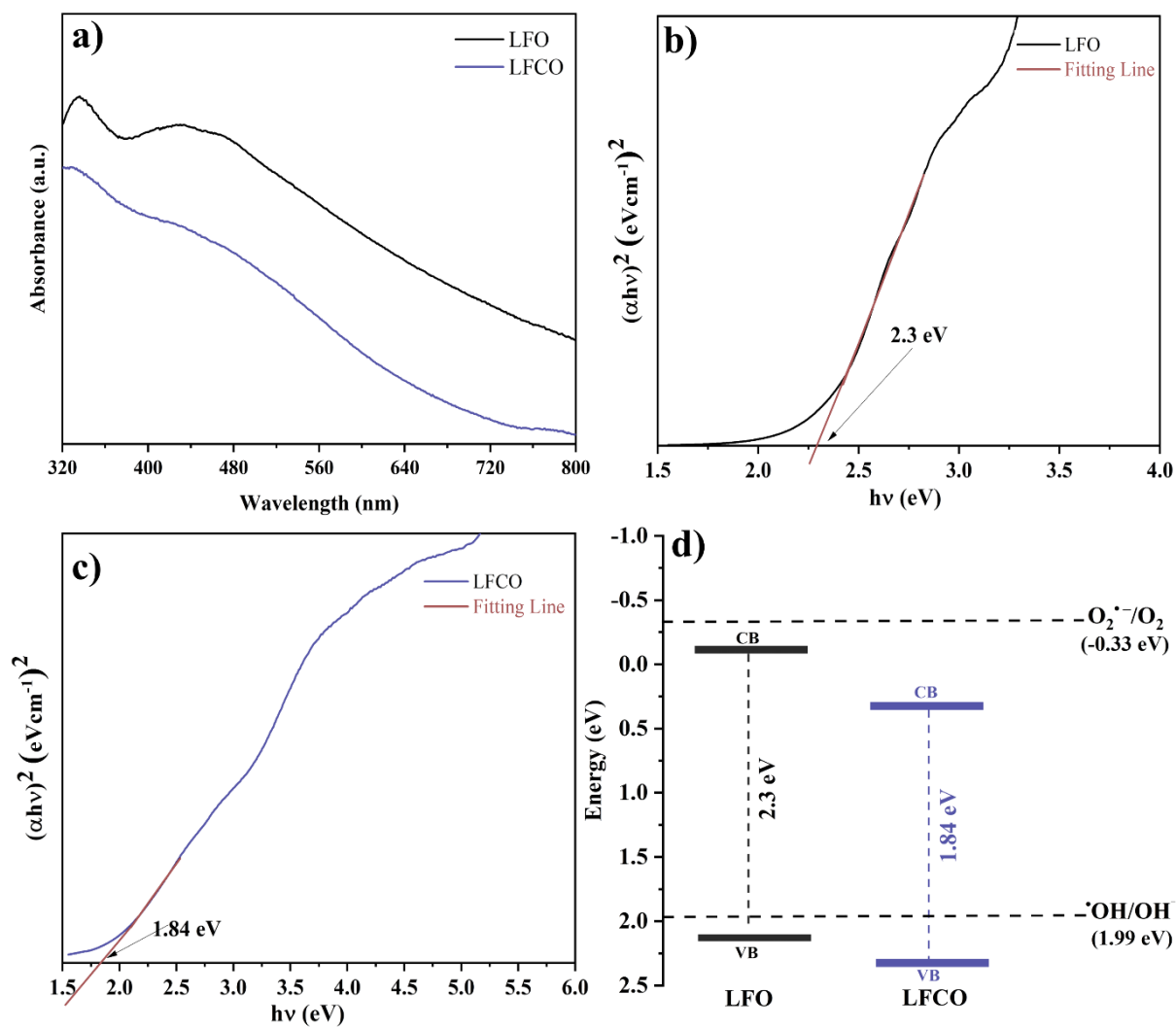

Fig. 5. a) UV-vis DRS spectra of LFO and LFCO samples, Tauc plot transformation of diffuse reflectance spectra of b) LFO, c) LFCO, d) schematic representations of the calculated Band structure (CB and VB) of LFO and LFCO NPs

### 3.3. MO dye degradation results:

To achieve adsorption-desorption equilibrium before starting the catalytic tests under visible light illumination and/or ultrasonic vibration, it is crucial to consider the zeta potential, which indicates the surface charge characteristics of LaFeO$_3$ and 10%Cu-doped LaFeO$_3$. The zeta potential measurements reveal that pure LFO has a surface charge of −14.4 mV, while LFCO exhibits a more negative surface charge of −24 mV. Note that the introduction of $Cu^{2+}$ ions into LFO promoted the formation of oxygen vacancies which can act as electron donors, leading to a higher concentration of negative charge at the surface, influencing the material's electronic structure and activity[74]. In addition, the increased charge negativity in the LFCO can enhance electrostatic repulsion against anionic dyes, such as Methyl Orange (MO), thereby avoiding the strong adsorption capacity on the LFCO surface.

Fig. 6 displays the results of the evaluation of the catalytic performances of LFO and LFCO NPs using MO degradation in water with different external environments. In this work, the absorbance of the initial MO (5 mg L$^{-1}$) was set as $C_0$ in Fig. 6(a), (d), and (g) for the three external stimuli photocatalysis, sonocatalysis, and sonophotocatalysis, respectively. The kinetic numbers in Fig. 6(c), (f), and (i) were the slopes of the fitted lines ($-\ln(C_t/C_0)$) in Fig. 6(b), (e), and (h).

As shown in Figs. 6(a), (d), and (g), after 120 minutes of stirring in the dark, the doped sample's adsorption is significantly lower than that of the pure LFO. These findings confirm our aforementioned results and align with previous research, which indicates that greater negative surface charges on catalyst materials lead to decreased adsorption of anionic dyes due to large electrostatic repulsion[55,75].

To assess the photocatalytic, sonocatalytic, and sonophotocatalytic performances of the synthesized samples, the decomposition ratio of MO dye in an aqueous solution was evaluated under the three external stimuli with mechanical stirring at 100 rpm as illustrated in Figures 6(a), (d), and (g). In the absence of a catalyst (Blank tests), the MO dye decomposition ratios are very slight (Fig. 6 (a), (d), and (g)).

The photocatalytic properties of LFO and LFCO NPs were evaluated under low-energy visible light (30 W). As shown in Fig. 6(a), the degradation of MO was very low, with only ~2–3% after 2 hours of visible-light photocatalysis. This limited photocatalytic activity may be attributed to the fast charge recombination, a common issue in $ABO_3$ perovskite oxides[76,77]. Previous studies highlighted this challenge, noting that the high rate of recombination in $LaFeO_3$ reduces its photocatalytic efficiency[13,42,46]. Meanwhile, several studies [42,78,79] showed that the addition of hydrogen peroxide ($H_2O_2$) increases the photodegradation rate of organic pollutants by removing surface-trapped electrons (Eq. 8), thereby lowering the electron-hole recombination rate and improving the efficiency of hole utilization for reactions such as (Eq. 9)[80]

$$LFO + h\nu \rightarrow e^-_{(CB)} + h^+_{(VB)} \quad Eq.\ 7$$

$$H_2O_2 + e^- \rightarrow {}^\cdot OH + OH^- \quad Eq.\ 8$$

$$OH^- + h^+ \rightarrow {}^\cdot OH \quad Eq.\ 9$$

$$H_2O_2 + h^+ \rightarrow {}^\cdot O_2^- + 2H^+ \quad Eq.\ 10$$

Interestingly, the addition of $H_2O_2$ markedly improved the photocatalytic performance of LFO and LFCO, as shown in the inset of Fig. 6(a). The enhancement is due to the generation of ${}^\cdot OH$ radicals through the reaction of $H_2O_2$ with photogenerated conduction band electrons ($e^-_{(CB)}$), which act as electron scavengers, thus inhibiting the recombination of $e^-/h^+$ pairs at the surface, according to equations 8 and 9[46]. Finally, the generated ${}^\cdot OH$ directly oxidized MO, as shown in Equation 11.

$$\text{·OH + MO} \rightarrow \text{Degradation products} \quad Eq.\ 2$$

These findings confirm that while LFO and LFCO NPs exhibit photo-responsive behavior under visible light, their efficiency is likely limited by electron-hole recombination.

The sonocatalytic performance of LFO and LFCO NPs was evaluated for the degradation of MO dye using low-energy ultrasonic vibrations (37 kHz, 300 W). As shown in Fig. 6(d), the results reveal significant differences in MO degradation efficiencies. In a control experiment without any catalyst (sonolyse), only 5% of MO degraded after 120 minutes, indicating that ultrasonic vibrations alone have a minimal effect on breaking down the dye. This suggests that acoustic cavitation, the process of bubble formation and collapse caused by ultrasound, plays a limited role in MO degradation without catalyst[81]. When LFO NPs were introduced, MO degradation improved significantly, achieving 93% degradation within 120 minutes. This enhanced performance is attributed to the excitation of LFO NPs by ultrasonic waves, which generate electron-hole pairs ($e^-/h^+$). According to the literature, the sonocatalytic degradation of organic pollutants in the presence of a catalyst can be explained by considering following three mechanisms: hot spot[19,82], sonoluminescence[14,21,35], and oxygen atom escape[21,83]. In the condition of only ultrasound, acoustic cavitation occurs as gas and vapor bubbles form, grow, and collapse violently within the liquid medium, creating localized hot spots with extremely high temperatures and pressures[21,84,85]. The sonogenerated electrons can react with electron acceptors such as $O_2$ dissolved in water and produce the superoxide anion radical ($·O_2^-$) (Eq. (12))[19]. Simultaneously, sonogenerated holes react with surface hydroxyl ions and water to form hydroxyl radicals (·OH) (Eqs. (9 and 13))[19].

$$e^- + O_2 \rightarrow {}^{\bullet}O_2^- \quad Eq.\ 3$$

$$H_2O + h^+ \rightarrow {}^{\bullet}OH + H^+ \quad Eq.\ 13$$

These reactive species are the primary agents responsible for dye degradation. Ultrasonication also prevents nanoparticle agglomeration, increasing the accessible surface area for catalytic reactions. The rapid collapse of cavitation bubbles removes surface contaminants, exposing fresh active sites and enhancing mass transport, which improves reactant diffusion to the catalyst surface. Moreover, under ultrasonic waves, the catalyst may act as a nucleation site for cavitation bubbles, amplifying radical production[21,86,87].

In contrast, Cu-doped LFO achieved only 55% degradation under similar conditions. This unexpected reduction in performance may result from introducing additional energy levels within the bandgap due to doping with transition metal ions like $Co^{2+}$, $Ni^{2+}$, $Zn^{2+}$, etc.[17]. While doping is generally intended to improve performance by adjusting the material's energy structure, it may inadvertently hinder sonocatalysis if the new energy levels are unsuitable for activation under ultrasound[87,88]. In addition, oxygen vacancies are known to induce a slight deterioration in sonocatalytic performance[88]. Similar effects have been observed in other doped materials, such as ZnO doped with W or Rh[15], as well as Dy-doped CdSe NPs[87] under sonocatalytic process.

Interestingly, combining light and ultrasonic excitations enhances the redox reactions during the sonophotocatalytic process. This synergistic effect arises from the generation of additional photogenerated electron-hole pairs ($e^-/h^+$), which are effectively separated by the mechanically formed fluctuating "wave" created by ultrasonic vibrations[18].

The MO dye degradation in the synergetic interaction of ultrasonic vibration and visible light illumination was utilized to assess the sonophotocatalytic efficacy of LFO and LFCO NPs. Interestingly, when LED-lamp irradiation and ultrasonication were simultaneously provided,

LFO and LFCO demonstrated superior catalytic ability compared to individual light irradiation or ultrasonic stimulation, as illustrated in Figure 6(g). Cu-doped LFO demonstrated the rapid and highest sonophotocatalytic activity, degrading MO concentration to approximately 100% within 120 minutes, while undoped LFO achieved 93% under simultaneous simulation. The coupling of ultrasound vibration with visible-light irradiation revealed a significant positive synergy between photocatalytic and sonocatalytic processes[89]. The results substantiated that the shockwaves from cavitation physically suppress the photo-generated ($e^-/h^+$) charges recombination, prolonging their lifetime, and producing additional hydroxyl groups. In addition, bubble collapse improves catalyst activation by refreshing active sites for photocatalysis, enhances light absorption by creating transient defects and breaking up particle clusters and improves the adsorption of the pollutant onto a fresh catalyst surface. The incorporation of $Cu^{2+}$ ions into LFO significantly enhances its sonophotocatalytic performance compared to undoped LFO under identical irradiation conditions. This enhancement can also be attributed to $Cu^{2+}$ doping-induced OVs, which serve as critical active sites optimizing charge carrier dynamics and light absorption, and mass transport[90]. These findings are supported by studies such as that of Wang et al.,[91] which demonstrates that OVs-rich ZnO enhances degradation efficiency through band gap narrowing and carrier trapping, and Wang et al.[92], showing that surface OVs on $WO_3$ promote charge separation by modifying the valence band and restricting electron-hole recombination. Additionally, L. Li et al.[93] found that OVs introduced into $AgNbO_3$ by $N_2$-annealing enhance piezophotocatalytic performance by improving absorbance and promoting electron-hole pair separation under an electric field. Our findings clearly demonstrate that Cu-doped LFO exhibits higher sonophotocatalytic degradation of MO compared to undoped LFO, making LFCO an excellent material for dye degradation by the sonophotocatalysis process.

To conduct a more methodical comparison of the sonophotocatalytic reaction rates (k) of the samples, a pseudo-first-order kinetic model, defined by Eq. (14), is used in Fig. 6(b),(e), and (h) and[56]:

$$-ln\left(\frac{c_t}{c_0}\right) = k.t \quad Eq.\ 44$$

where k (min$^{-1}$) and t (min) are the reaction rate constant and the catalytic time, respectively. When using combined processes, the kinetic analysis shows significant improvements in reaction rates for both undoped and Cu-doped LFO samples (Fig. 6(h)). For undoped LFO, the sonophotocatalytic process yields a rate constant of 0.021 min$^{-1}$, significantly higher than that of photocatalysis (0.00048 min$^{-1}$) or sonocatalysis (0.013 min$^{-1}$) alone. More remarkable results are demonstrated by Cu-doped LFO, which exhibits rate constants of 0.00053, 0.0042, and 0.0455 min$^{-1}$ for photocatalytic, sonocatalytic, and sonophotocatalytic processes, respectively. This substantial enhancement in reaction rate for Cu-doped LFO underscores the dual benefits of Cu doping and process combination.

The synergistic effect of a sonophotocatalysis process can be quantified using the synergistic index (SI)[94], calculated as the ratio of the rate constant of sonophotocatalysis to the sum of the rate constants of the individual processes (Eq. 15)[94,95],

$$SI = \frac{k(sonophotocatalysis)}{k(sonocatalysis) + k(photocatalysis)} \quad Eq.\ 5$$

This index is commonly employed to analyze the degree of synergistic enhancement in dye degradation. A SI value greater than 1 indicates that the efficiency of sonophotocatalytic degradation surpasses the combined effect of individual processes, demonstrating a positive synergistic effect[94]. A high synergistic index of about 9.62 was established by the Cu-doped LFO sample under sonophotocatalysis. This finding is consistent with several studies in the literature that have demonstrated the enhanced efficiency of organic pollutant degradation through the combined sonophotocatalytic approach. Benomara et al.[27] reported pseudo-first-order rate constants for methyl violet 2B degradation: $6.8 \times 10^{-3}$ (sonocatalysis), $22.9 \times 10^{-3}$ (photocatalysis), and $39.7 \times 10^{-3}$ min$^{-1}$ (sonophotocatalysis), indicating a significant synergistic effect of using the sonophotocatalysis process. Similarly, Babu et al.[96] achieved a synergistic index of 3.7 for Methyl orange degradation using CuO-TiO$_2$/rGO nanocatalysts in sonophotocatalysis. Ahmad et al.[97] investigated Rhodamine B (RhB) degradation using 10 wt.% CNTs–ZnO, reporting rate constants of $10 \times 10^{-3}$ min$^{-1}$ (photocatalytic), $11 \times 10^{-3}$ min$^{-1}$ (sonocatalytic), and $44 \times 10^{-3}$ min$^{-1}$ (sonophotocatalytic), with the sonophotocatalytic process exhibiting a higher rate constant than the sum of individual processes (SI=2.09). The results obtained from the current study were compared with those previously reported for sonophotocatalytic degradation of dyes (Table 2), demonstrating the high efficiency of LFCO as a dye degradation material.

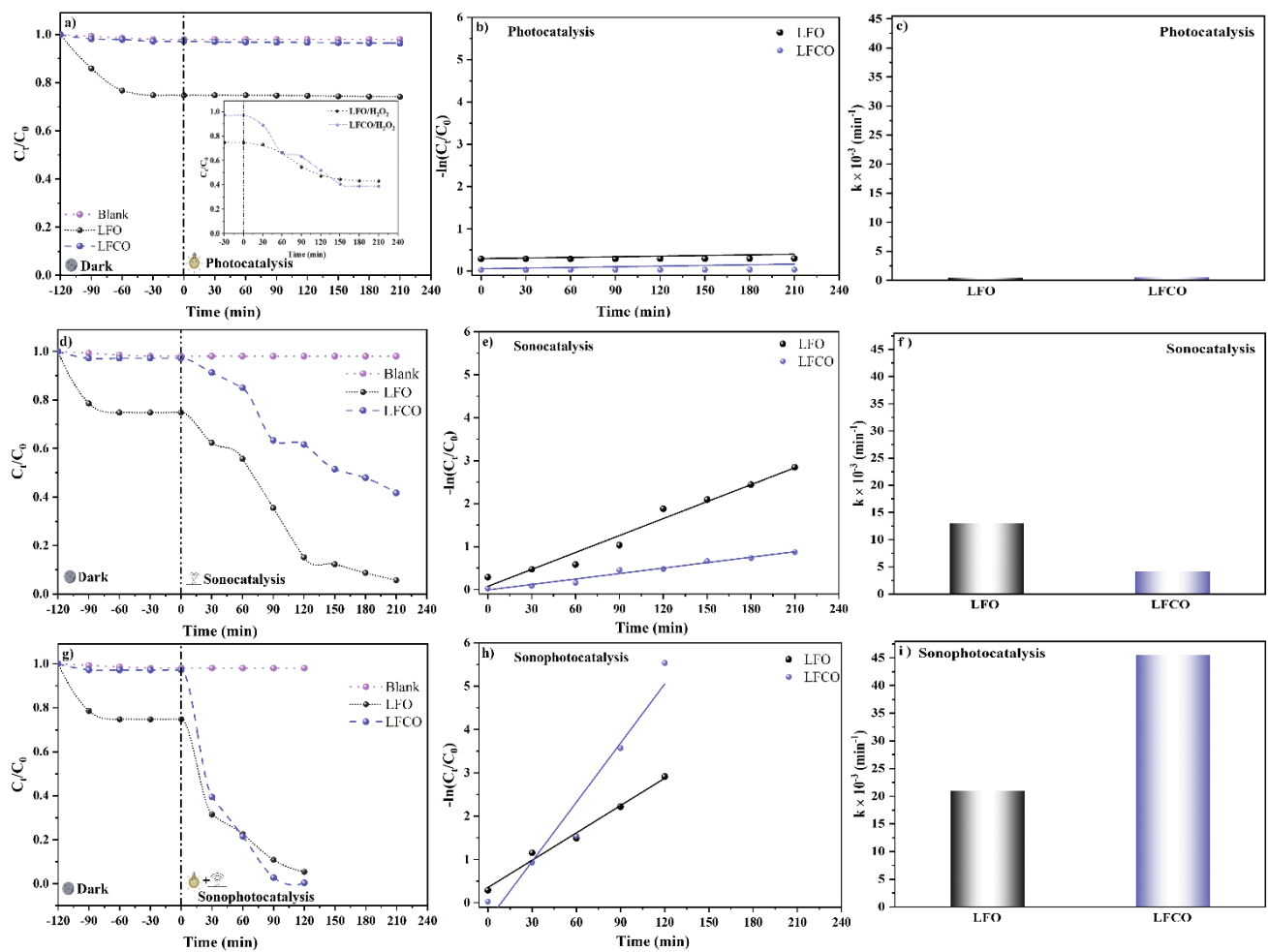

Fig.6. MO degradation curves ((a), (d), (g)), kinetic curves ((b), (e), (h)), and comparison of degradation ability ((c), (f), (i)) of LFO and LFCO NPs in different external environments.

Table 2. Comparison of sonophotocatalytic activity of LFO and LFCO NPs for dye degradation with other reported catalysts.

| Catalysts | Energy source (U-ultrasonic+L-light) | Dye pollutant | Degradation efficiency (%) | Kinetic rate constants ($\times 10^{-3}$ min$^{-1}$) | Synergetic index | Refs |
|---|---|---|---|---|---|---|
| LFCO NPs | U(37 kHz/300 W)+ L( LED 30W) | MO 5mg L$^{-1}$ | 100%;120 mn | 45.5 | 9.61 | This work |
| LFO NPs | U(37 kHz, 300 W)+L( LED 30W) | MO 5mg L$^{-1}$ | 93%;120 mn | 21 | 1.56 | This work |
| La$_{0.8}$FO@PgNS nanosheet | U(Not specified)+ L(Xenon lamp 300W) | Sulfamethoxazole 5mg L$^{-1}$ | 99.6%;120 mn | 14.9 | _ | 16 |
| 1% Er-doped BVO/BFO nanocomposites | U(200 W, 20 kHz)+L(Xe lamp 300W ) | RhB - | 99% ; 90 mn | 48.4 | 2.11[a] | 98 |
| BiFeO$_3$ nanowires | U(80 W, 132 kHz)+L (Xe lamp 300W ) | RhB 4 mg L$^{-1}$ | 97%; 60 mn | 58.2 | 1.16[a] | 99 |
| Ag modified Na$_{0.5}$Bi$_{0.5}$TiO$_3$ nanospheres | U(40KHz,50W) + L(UV light Xenon lamp, 300W) | RhB 5 mg L$^{-1}$ | 98.6%; 120 mn | 32.9 | 6.7[a] | 100 |

| BaTiO$_3$ nanowires | U (40 kHz, 180 W)+L (UV-LED) | MO 5 mg L$^{-1}$ | 98.1%; 80 mn | 42.63 | 1.6$^a$ | 101 |
| Na$_{0.5}$Bi$_{2.5}$Nb$_2$O$_9$ nanosheets | U (600 W, 40 kHz) + L (Xe lamp 300 W) | RhB 20 mg L$^{-1}$ | 98%; 120 mn | 24.8 | – | 102 |

$^a$ The value was derived from calculations based on data presented in the cited articles

Furthermore, the sonophotocatalytic degradation efficiency of LFO and LFCO was evaluated by measuring the Total Organic Carbon (TOC) values of the MO working solution before and after the experiment. This assessment provides insight into the degree of mineralization of contaminants into $CO_2$ and $H_2O$. The sonocatalysis shows a TOC removal higher than 50 % (Fig. 7(a)), while using the sonophotocatalytic process, the TOC removal percentage was observed to be nearly 90% (Fig. 7(b)), indicating substantial mineralization of the MO dye. Notably, this value closely aligns with the 100% MO degradation observed through UV-VIS spectrophotometry, indicating a good agreement between the two evaluations.

The catalyst must demonstrate not only good efficiency and versatility but also high reusability and stability so that it can be recycled and reused. Reusability is a critical factor influencing the application of nanomaterials in water purification and their economic viability in wastewater treatment. The reusability of LFCO for sonophotocatalytic degradation of MO was evaluated over four cycles. After each cycle, the catalyst was recovered by centrifugation, washed with distilled water and absolute ethanol, dried at 60 °C, and reintroduced to a fresh MO solution for the next cycle under identical conditions. The experiments were conducted under simultaneous irradiation with a low-energy LED lamp and ultrasonication for 2 hours. Fig. 7(c) illustrates the reusability of LFCO for MO sonophotocatalytic elimination over four cycles. A gradual decrease in catalytic performance was observed, with approximately 28% reduction by the fourth cycle. Notably, minimal performance decline was observed between the second and third cycles. Despite this gradual decrease in efficiency, LFCO NPs retained considerable sonophotocatalytic ability even after prolonged LED-lamp irradiation and ultrasonic vibration. This performance is comparable to, or better than, many other reported catalysts in terms of reusability[99]. The slight decrease in efficiency could be attributed to minimal catalyst loss during the recycling process, which is a common challenge in heterogeneous catalysis. In addition to reusability, structural stability is crucial for promoting technological applications in sonophotocatalysis. To assess this, XRD analysis was performed for LFCO NPs before and after four sonophotocatalytic cycles (Fig. 7(d)). Both XRD patterns exhibited similar diffraction peaks with no detectable secondary phases, demonstrating excellent structural stability. These findings demonstrate the good reusability of Cu-doped LFO, positioning it as a promising candidate for efficient and sustainable sonophotocatalytic applications.

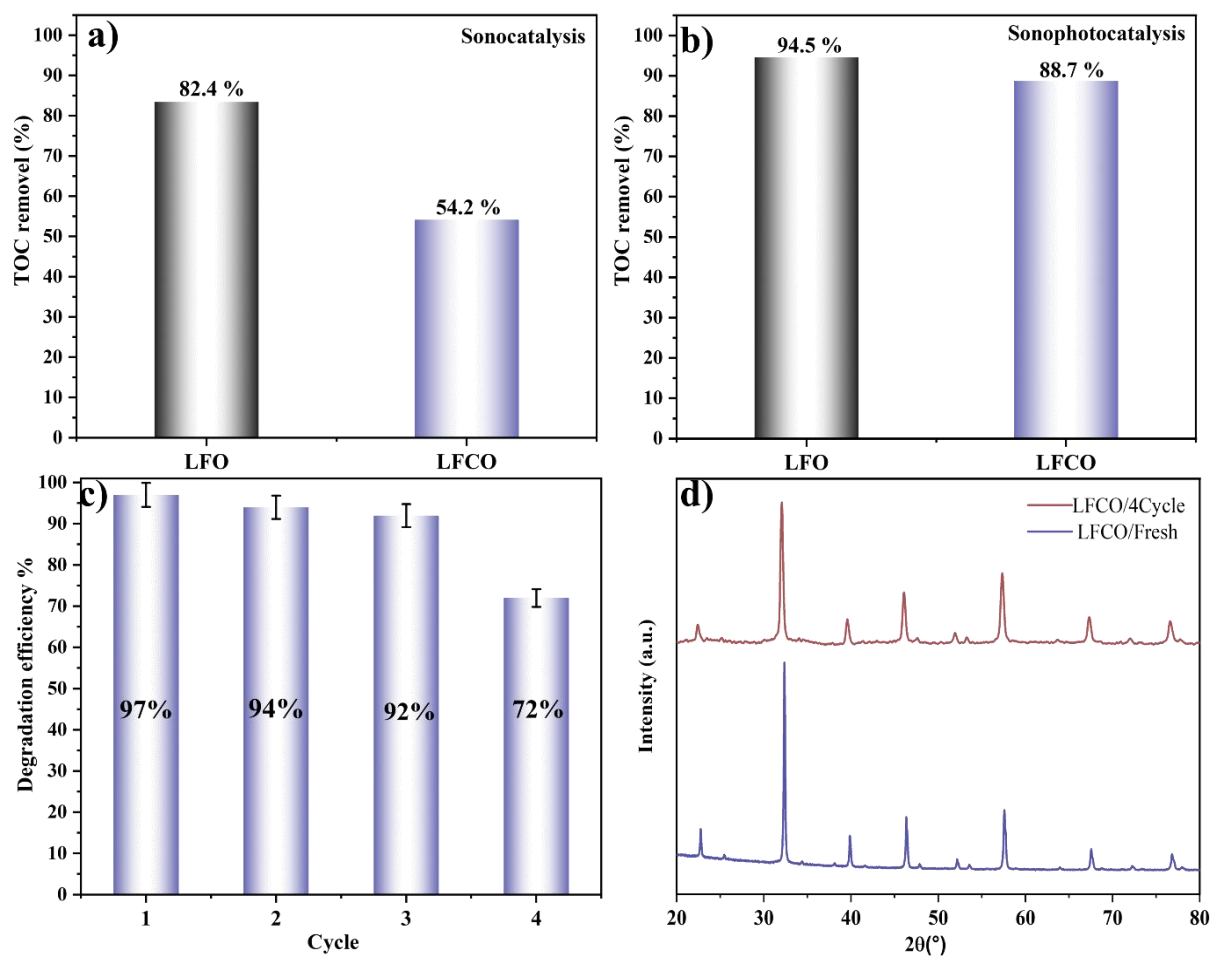

Fig. 7. Proportion of TOC removal analyzed upon MO a) sonocatalytic, b) sonophotocatalytic degradation using LFCO NPs, c) Cyclic sonophotocatalytic degradation of MO over LFCO, d) XRD patterns of LFCO before and after 4-cycle degradation experiments.

Based on the above results, a possible sonophotocatalytic degradation pathway for methyl orange in the LFCO system is proposed, as illustrated in Fig. 8(a). The integration of photocatalysis and sonocatalysis ensures the efficient utilization of reactive oxygen species (ROS) on LFCO catalysts, while maintaining high degradation efficiency for persistent organic pollutants, such as MO[26]. The combination of photocatalytic and sonocatalytic technologies results in several advantages: (i) the generation of multiple ROSs through both photocatalytic reactions and cavitation effects[82], (ii) increased availability of active sites due to catalyst particle disagglomeration by ultrasound[103], (iii) improved mass transfer of MO molecules to the catalyst surface via cavitation effects[26], and (iv) the formation of hot spots that facilitate localized high-energy reactions[3]. To fully understand the sonophotocatalytic mechanism, it is essential to examine each process individually. The sonolysis of water induced by ultrasound generates cavitation bubbles that collapse under extreme conditions of high temperature and pressure[14]. This collapse can lead to the pyrolytic breakdown of water molecules into hydroxyl radicals (•OH) and hydrogen radicals (H•), as shown in the following reactions[14,16]:

LFCO + hν → LFCO (e⁻ + h⁺)     *Eq. 16*

LFCO + ultrasound ))))→ LFCO (e⁻+h⁺) *Eq. 6*

H₂O + ultrasound ))))→ H• + •OH    *Eq. 7*

•OH + •OH → H₂O₂    *Eq. 19*

H₂O₂ + e⁻ → •OH + OH⁻   *Eq.20*

H• + O₂ → HO₂•    *Eq. 21*

HO₂ + HO₂ → O₂ + H₂O₂ → 2 •OH   *Eq. 8*

H₂O₂ + •O₂⁻ → •OH + OH⁻ + O₂   *Eq. 23*

h⁺ + MO → degraded products    *Eq. 24*

Simultaneously, under light irradiation, the semiconductor catalyst LFCO absorbs photons with energy equal to or greater than its band gap[21]. This excitation promotes electrons (e⁻) from the VB to the CB, leaving behind holes (h⁺) in the VB. These charge carriers drive redox reactions on the catalyst surface (Eq. 16). Holes oxidize water or hydroxide ions to generate additional hydroxyl radicals (•OH) (Eq. 13)[12]. Thus, the catalytic redox reaction is enhanced, thereby accelerating the efficiency of dye degradation. These ROS also contribute to the innocuous end products formed as MO dye degrades (Eq. 11)[104].

Hydroxyl radicals ($^\bullet$OH), in particular, attack the azo bond (−N=N−) in MO molecules, leading to its breakdown into intermediate products such as N, N-dimethyl-p-phenylenediamine and N, N-dimethyl-benzenamine[104]. These intermediates are further oxidized into simpler compounds such as carbon dioxide ($CO_2$), and water ($H_2O$)[86]. This proposed mechanism aligns with previous studies that emphasize the dual role of ultrasound in enhancing both physical processes (e.g., mass transfer and particle disagglomeration) and chemical processes (e.g., ROS generation)[32,104]. The synergy between photocatalysis and sonocatalysis lies in their complementary mechanisms, which enhance the generation of ROS and improve the particles dispersion and the sonophotocatalytic degradation efficiency of organic pollutants. This combined approach not only accelerates the degradation process but also potentially reduces the formation of potentially toxic intermediates, making it a promising method for wastewater treatment and environmental remediation.

To confirm the above proposed mechanism, where $^\bullet$OH, $h^+$, and $^\bullet O_2^-$ radicals are the primary active species for organic dye degradation, radical-trapping experiments were conducted using different scavengers. The introduction of these scavengers significantly reduced the decomposition rates compared to reactions without scavengers, as the scavenging agents captured a portion of the active species, thereby limiting sonophotocatalytic activity. As shown in Fig. 8b, adding 1 mM isopropanol alcohol (IPA, an $^\bullet$OH scavenger) substantially inhibited the catalytic process, resulting in only 6% MO dye degradation within 120 min . When 1 mM ethylenediaminetetraacetate (EDTA, a $h^+$ scavenger) was introduced, the catalytic process was also reduced  to 60% of the dye degraded. In contrast, the addition of $N_2$ (an $O_2^{\bullet -}$ scavenger) had a less pronounced effect, allowing for 80% MO dye degradation. These results indicate the dominant role of holes ($h^+$) and $^\bullet$OH radicals in the sonophotocatalytic degradation of MO as predicted from the energy band levels of the two catalysts (Fig. 5d). The minimal impact of $^\bullet O_2^-$ radicals, which electrons cannot produce, further confirms this conclusion as shown in Fig.

(8, & 5d). Indeed, at pH≈ 6, the conduction band minimum of LCFO (0.47eV) is more positive than the oxygen reduction reaction potential (E°$_{RHE}$= -0.33 eV). Whereas, the LCFO conduction band level (E$_{VB}$ = 2.31 eV) is much higher positive than the conduction band level of the water oxidation redox couple (E°= 1.99 eV vs. RHE), exhibiting enhanced efficiency in degrading methyl orange dye by preferentially generating hydroxyl radicals (•OH) over superoxide. This scavenger study provides strong evidence supporting the proposed mechanism and highlights the importance of hydroxyl radicals (•OH) and holes (h$^+$) in the degradation process.

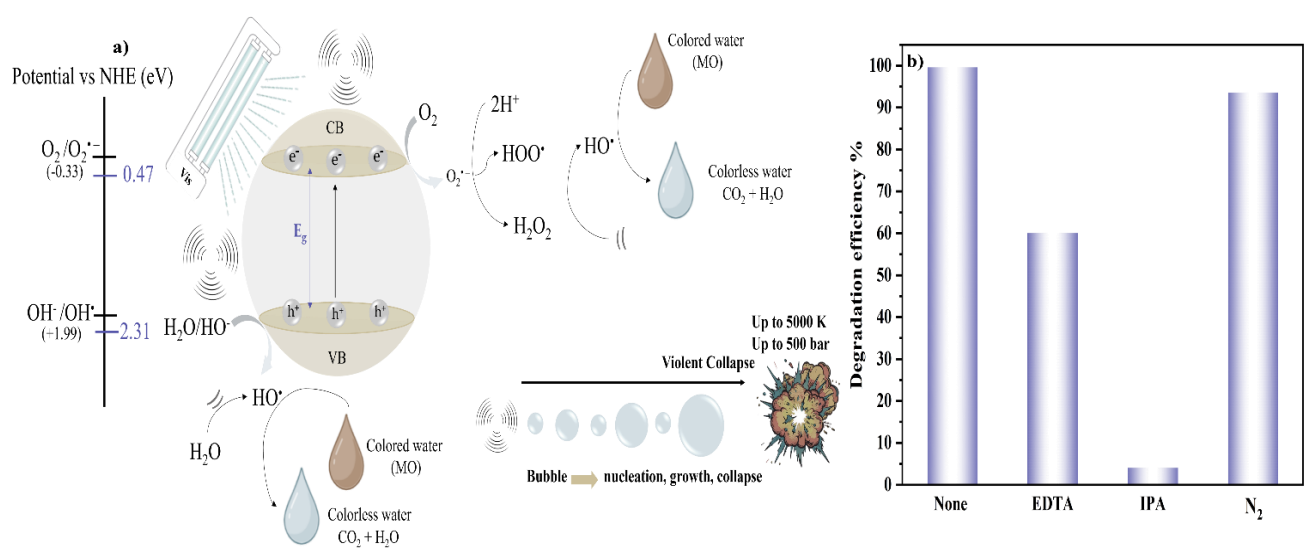

Fig. 8. a) Proposed sonophotocatalytic mechanism and possible pathway for degradation of MO by LFCO NPs, b) sonophotocatalytic degradation of MO by LFCO in the presence of scavengers

**Conclusion**

In this study, the catalytic degradation performance of LaFeO$_3$ (LFO) and Cu-doped LaFeO$_3$ (LFCO) nanoparticles as catalysts was comprehensively investigated through the degradation of MO using photo, sono, and sonophotocatalytic oxidation under visible light and low-frequency ultrasound. Both undoped LFO and LFCO NPs were successfully synthesized using the sol-gel auto-combustion method. Comprehensive structural and morphological analyses were conducted using XRD, Raman spectroscopy, TEM, EDX, and UV-Vis techniques. XRD patterns confirmed the formation of a pure orthorhombic structure for both undoped and Cu-doped LFO NPs annealed at 900 °C, with no secondary phases detected. Raman spectroscopy further validated the absence of copper oxide formation upon Cu ion incorporation into the LFO lattice. TEM revealed that the synthesized NPs exhibited random, sphere-like morphologies, forming clusters. Cu doping generated oxygen vacancies, inducing contraction of the perovskite structure. The Cu-doped LFO NPs exhibited outstanding performance, achieving complete degradation of the MO dye within 2 hours under synergistic sonophotocatalytic conditions with a high kinetic rate constant of $45.5 \times 10^{-3}$ min$^{-1}$. The rapid and more effective MO removal observed under visible irradiation and ultrasonic vibration was attributed to the synergistic effect of photocatalysis and sonocatalysis stimulus. Trapping experiments identified holes and $^{\bullet}$OH radicals as the primary reactive species driving the degradation mechanism. Additionally, reusability tests demonstrated excellent stability, with the catalyst retaining over 70% of its activity after multiple cycles and showing strong resistance to photo-corrosion. These findings position Cu-doped LFO NPs as highly efficient and durable catalysts with significant potential for advanced environmental applications, including pollutant degradation, hydrogen production, and methane generation.

## Acknowledgements

This work was supported by the Horizon Europe Framework Program Action HORIZON-MSCA-2022-SE-01-H-GREEN (No. 101130520), the Region of Hauts-De-France (HDF), and the Slovenian Research Agency (project J2-60035, core funding P2-0105 and P1-0125). The technical assistance of Jena Cilenšek and Killian Guillou (Erasmus+) is greatly acknowledged.